\documentclass[english,nofootinbib]{revtex4}
\usepackage[T1]{fontenc}
\usepackage[latin9]{inputenc}
\usepackage{babel}
\usepackage{amstext}
\usepackage{graphicx}
\usepackage{esint}
\usepackage[unicode=true,pdfusetitle,
 bookmarks=true,bookmarksnumbered=false,bookmarksopen=false,
 breaklinks=false,pdfborder={0 0 1},backref=section,colorlinks=false]
 {hyperref}
\usepackage{breakurl}
\usepackage{amstext}\usepackage{breakurl}
\makeatletter

\def\beq{\begin{equation}}
\def\eeq{\end{equation}}
\def\bea{\begin{eqnarray}}
\def\eea{\end{eqnarray}}

\def\eq#1{{Eq.~(\ref{#1})}}

\makeatother

\begin{document}

\title{On relation between rest frame and light-front descriptions of quarkonium}

\author{B.~Z.~Kopeliovich, E. Levin, Iv\'an~Schmidt and M.~Siddikov}

\affiliation{Departamento de F\'{i}sica, Universidad T\'ecnica Federico Santa
Mar\'{i}a,\\
 y Centro Cient\'{i}fico - Tecnol\'ogico de Valpara\'{i}so, Casilla
110-V, Valpara\'{i}so, Chile}

\preprint{USM-TH-332}
\begin{abstract}
In this paper we study the relation between the light-front (infinite momentum) and rest-frame
descriptions of quarkonia. While the former is more convenient
for high-energy production, the latter is usually used for the evaluation
of charmonium properties. In particular, we discuss the dynamics of
a relativistically moving system with nonrelativistic internal motion
and give relations between rest frame and light-front potentials used
for the description of quarkonium states. We consider two approximations,
first the small coupling regime, and next the nonperturbative
small binding energy approximation. In both cases we get consistent
results. Our results could be relevant for the description of final
state interactions in a wide class of processes, including quarkonium
production on nuclei and plasma. Moreover, they can be extended to
the description of final state interactions in the production of weakly bound
systems, such as for example the deuteron. 
\end{abstract}
\maketitle

\section{Introduction}

In the general case, a direct boosting of the wave function components
from one frame to another presents a complicated dynamical problem,
which mixes states with various parton content~\cite{Lepage:1980fj}.
However, for certain systems, for example positronium, 
heavy quarks systems, or a nucleus with vanishingly small
binding energy, the Fock state is dominated by the lowest component
with minimal number of partons. The dynamics of such systems is described
in potential models, and the wave functions in different reference frames
can be related to each other: it is well-known that in the light front
quantization approach the wave function is invariant with respect to longitudinal boosts~\cite{Lepage:1980fj,Brodsky:2009cf,Brodsky:1997de},
so a transformation of the rest frame wave function requires just
to rewrite it in terms of light-front variables ~\cite{Ivanov:2002kc}.
This solves a boosting problem for a class of processes, in which
due to factorization theorems only the light front wave function of
the initial or final hadron is needed. However, in certain cases a more 
sophisticated approach is necessary: when, for example, processes inside extended
objects like nuclei~\cite{Kopeliovich:2001ee,Kopeliovich:2002yv}
and plasmas~\cite{Kopeliovich:2014uha,Kopeliovich:2014una} are considered.
In the evolution of such systems, an effective potential should be taken
into account at all stages. In the light front, which is a natural
choice for description of high energy processes, usually the quark
potentials are much less understood than in the rest frame. For example,
nothing is known about the light-front potentials at nonzero temperatures.
The main goal of this manuscript is to fill this gap and provide a
method which would bridge the light front and rest frame approaches.
For this purpose we use the relation between the light front and the so called infinite momentum frame.

In what follows, for the sake of definiteness we will consider the particular
case of quarkonium (charmonium or bottomonium), tacitly assuming that
results could be extended to other systems. Quarkonium has been
very well understood in the rest frame~\cite{Korner:1991kf,Brambilla:1999xf,Arafah:1983,Jacobs:1986,Gara:1990,Ikhdair:1992,Brambilla:1992,Fulcher:1993,Oh:2002,Blanck:2011,Beneke:2005hg,Brambilla:2004wf,Brambilla:2010vq},
even for nonzero temperature~\cite{Matsui:1986dk,Digal:2005ht,Eichten:1979ms,Kaczmarek:2003ph,Kaczmarek:2004gv,Kaczmarek:2012ne,Karsch:1987pv,Karsch:2005nk}.
Experimentally there are data for quarkonia production both on
the proton and nuclear targets (see e.g.~\cite{Bedjidian:2004gd,Brambilla:2004wf,Kopeliovich:2010jf}
for review and references therein). In photo- and electroproduction
on protons, factorization theorems hold, and as was mentioned
above, the final state wave function can be obtained ~\cite{Ivanov:2002kc}.
In production on nuclear targets, as was discussed in~\cite{Kopeliovich:2001ee,Kopeliovich:2002yv},
the contribution of final state interactions (FSI) is important
and requires knowledge of the light-front potentials. Recently the
relation between the rest frame and light front quarkonium potentials
has been studied in AdS/QCD framework~\cite{Trawinski:2014msa,Gutsche:2014oua}.
Basing on equality of the spectra, it was suggested that the transformation
could be nonlinear, albeit inclusion of the higher order $\mathcal{O}(1/m_{q})$
terms inevitably has some ambiguity and differs between different
authors~\cite{Trawinski:2014msa,Gutsche:2014oua}. From our point
of view, inclusion of such terms is not justified since at the same
order there are contributions from the omitted multiparton Fock states.

In this paper we re-visit the problem of bound state of two heavy quarks, where the whole system is moving
with relativistic momentum in the laboratory frame. In Section~\ref{sec:Comoving}
we analyze the dynamics of the system perturbatively in the moving
frame and reduce a Bethe-Salpeter equation to a Schroedinger equation~(\ref{eq:Schroedinger_JPSi})
for a nonrelativistic internal motion of a moving dipole. The latter
is preferable for many practical applications because the near-onshellness
of both heavy quarks in a Bethe-Salpeter equation complicates the numerical
treatment~\cite{Carbonell:2014dwa}. The novelty of our approach
is that, in contrast to existing treatments, from the very beginning
we consider a moving quarkonium. Also, we give a generalization of
the pNRQCD lagrangian~(\ref{eq:pNRQCDLagrangian}) in a moving frame.

In view of the fact that for charmonium $\alpha_{s}\left(m_{c}\right)\approx0.25$
is not very small, in Section~\ref{sec:AnalyticitySchroedinger}
we develop a different approach to the non-relativistic system, based
on analyticity and unitarity constraints, in which we assume  smallness
of the parameter $\epsilon/\Lambda_{QCD}$ and which is not based (at
least explicitly) on the smallness of the QCD coupling. In this approach,
we get exactly the same Schroedinger equation~(\ref{eq:Schroedinger_JPSi}).
In this method of derivation we pursue a practical aim: to formulate
a set of rules on how to use the phenomenology developed for the description
of $J/\Psi$-meson in the laboratory frame, in any moving reference
frame.

\section{Schroedinger equation for heavy quarkonium in a moving
reference frame, from the Bethe-Salpeter equation}

\label{sec:Comoving}

As it is well known~\cite{Korner:1991kf,Brambilla:1999xf}, in the
rest frame there is a natural small parameter, the velocity of the heavy
quark $v\ll1$. A systematic expansion over this parameter leads to
an effective theory, NRQCD. In a perturbative pNRQCD, (which requires
an additional assumption $\alpha_{s}(m_{q})\ll1$) the velocity of
internal motion of the bound quark $v\sim\alpha_{s}$, so a small-$v$
expansion is equivalent to a systematic perturbative expansion. For
asymptotically heavy quarks, there is a hierarchy of scales 
\begin{equation}
m_{q}\gg m_{q}\alpha_{s}\gg m_{q}\alpha_{s}^{2}\gg\Lambda_{{\rm QCD}}\label{eq:Hierarchy-1-1}
\end{equation}
where $m_{q}$ is the mass of the heavy (charm or bottom) quark. The
scales in this hierarchy have a straightforward physical meaning:
the soft scale $m_{q}\alpha_{s}$ corresponds to inverse size of the
quarkonium system; the ultrasoft scale $m_{q}\alpha_{s}^{2}$ is the
typical binding energy~$\epsilon$, etc~\cite{Korner:1991kf,Brambilla:1999xf}.

\begin{figure}
\begin{centering}
\includegraphics[scale=0.5]{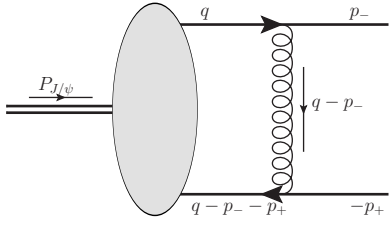} 
\par\end{centering}

\caption{Bethe-Salpeter kernel for the two-parton system.\label{fig:BS}}

\label{rf} 
\end{figure}

The dynamics of the system is described by the Bethe-Salpeter equation
(BSE), which is explicitly invariant in any system and contains both
a rest frame and a light front Schroedinger equations as special limits.
In an explicit form, the BSE is written as (see Fig.~\ref{rf}) 
\begin{equation}
i\chi_{ab}\left(p_{-},-p_{+}\right)=S_{ac}\left(p_{-}\right)\int\frac{d^{4}q}{(2\pi)^{4}}\tilde{A}_{ce,fd}\left(p_{-},q-p_{+}-p_{-},q-p_{+}\right)\,\chi_{ef}\left(q,\, q-p_{-}-p_{+}\right)S_{db}\left(-p_{+}\right),\label{eq:BSE_def-1}
\end{equation}
where $p_{-}$ and $-p_{+}$ are momenta of the quark and antiquark
respectively. 
In~\eq{eq:BSE_def-1} $\tilde{A}_{ce,fd}$ denotes the 2-quark scattering
amplitude. Recall that in this equation all ingredients are relativistic
invariants and do not depend on the reference frame.

In a perturbative QCD, the propagator $S$ in the leading order should
be replaced by the free quark propagator, and the vertex part $\tilde{A}$
should be replaced by a single-gluon exchange in the $t$-channel, so~(\ref{eq:BSE_def-1})
simplifies to 
\begin{equation}
i\chi\left(p_{-},-p_{+}\right)=S\left(p_{-}\right)\int\frac{d^{4}q}{(2\pi)^{4}}\frac{\gamma_{\mu}\chi\left(q,\, q-p_{-}-p_{+}\right)\gamma_{\mu}}{\left(q-p_{-}\right)^{2}+i0}S\left(-p_{+}\right),\label{eq:BSE_1g-2}
\end{equation}
where from now on we omit explicit Dirac indices. If we denote $p_{0}=P/2$,
where $P$ is the momentum of the quarkonium, and define small deviations
as $\delta p_{-}=p_{-}-p_{0},\quad-\delta p_{+}=-p_{+}-p_{0}$, after
some well-known algebraic manipulations (see details in Appendix~\ref{sec:BSE_derivation}),
we may reduce~(\ref{eq:BSE_1g-2}) to 
\begin{eqnarray}
i\tilde{\chi}\left(p_{-},-p_{+}\right) & = & g\left(\delta p_{-}\right)g\left(-\delta p_{+}\right)\int\frac{d^{4}q}{(2\pi)^{4}}\,\frac{\tilde{\chi}\left(q,\, q-p_{-}-p_{+}\right)}{\left(q-p_{-}\right)^{2}+i0},\label{eq:BSE_1g-1-1}\\
\tilde{\chi} & \equiv & \Lambda_{+}\gamma_{\mu}\chi\gamma_{\mu}\Lambda_{-}
\end{eqnarray}
Note that while $\tilde{\chi}$ is a matrix, equation~(\ref{eq:BSE_1g-1-1})
is a scalar equation (doesn't mix components), and for this reason in
what follows we may omit the spin structure of $\tilde{\chi}$ and
treat it as a scalar function. Taking into account that the projection
of the vector $q-p$ on $P$ is $\mathcal{O}\left(\alpha_{s}^{2}\right)$
(this corresponds to a generalization of the rest frame instantaneous
approximation $k_{0}\approx0$), after some algebra (see Appendix~\ref{sec:BSE_derivation})
we can obtain a Schroedinger equation for the internal motion in the
form 
\begin{equation}
\left(i\partial_{\xi}+\frac{\partial_{\zeta}^{2}}{m_{q}}-\frac{p_{\perp}^{2}}{m_{q}}\right)\Psi\left(\zeta,r_{\perp}\right)=\tilde{V}\left(\zeta,r_{\perp}\right)\Psi\left(\zeta,r_{\perp}\right),\label{eq:Schroedinger_JPSi}
\end{equation}
where we defined a wave function $\Psi$ as $\Psi=\int d\lambda\,\tilde{\chi}\left(q,\, q-P\right)_{_{\lambda=q\cdot P/M}},$
a potential $\tilde{V}$ is proportional to the trace of the gluon propagator,
$\tilde{V}=g^{\mu\nu}\Pi_{\mu\nu}/4$, where $\Pi_{\mu\nu}$ is the
gluon propagator, and the derivatives are defined as 
\begin{eqnarray}
\partial_{\xi} & = & -\left(\cosh\eta\partial_{0}+\sinh\eta\partial_{3}\right),\\
\partial_{\zeta} & = & -\left(\sinh\eta\partial_{0}+\cosh\eta\partial_{3}\right).
\end{eqnarray}
Here we introduced the shorthand notations $\eta=\ln\left(P^{+}/M\right)$,
and $M$ for the mass of quarkonium. The relation of the effective
potential $\tilde{V}$ to the rest frame potential will be discussed
in the next section. Equation~(\ref{eq:Schroedinger_JPSi}) provides
a smooth interpolation between the rest frame ($\eta=0$) and light-front
frame ($\eta\to\infty$). This equation can be also 
obtained from the pNRQCD lagrangian~\cite{Brambilla:1999xf},
which in a moving frame gets the form 
\begin{eqnarray}
L & = & \Psi_{S}^{\dagger}\left(i\partial_{\xi}+\frac{\partial_{\zeta}^{2}}{m_{q}}-\frac{p_{\perp}^{2}}{m_{q}}-V_{S}\right)\Psi_{S}+\Psi_{O}^{\dagger}\left(iD_{\xi}+\frac{\partial_{\zeta}^{2}}{m_{q}}-\frac{p_{\perp}^{2}}{m_{q}}-V_{O}\right)\Psi_{O}+\label{eq:pNRQCDLagrangian}\\
 & + & \frac{g}{m_{q}}\, V_{A}(r)\,\left(\Psi_{S}^{\dagger}r_{\perp}^{\mu}P^{\nu}F_{\mu\nu}\Psi_{O}+\Psi_{O}^{\dagger}r_{\perp}^{\mu}P^{\nu}F_{\mu\nu}\Psi_{S}\right)+\nonumber \\
 & + & g\,\frac{V_{B}(r)}{2m_{q}}\,\left(\Psi_{O}^{\dagger}r_{\perp}^{\mu}P^{\nu}F_{\mu\nu}\Psi_{O}+\Psi_{O}^{\dagger}r_{\perp}^{\mu}P^{\nu}F_{\mu\nu}\Psi_{O}\right)-\frac{1}{4}F^{a,\mu\nu}F_{\mu\nu}^{a},\nonumber 
\end{eqnarray}
where $\Psi_{S}$ and $\Psi_{O}=\Psi_{O}^{a}t_{a}$ are the fields
of the singlet and octet $\bar{q}q$-pair ($\text{\ensuremath{\Psi}}$
in~(\ref{eq:Schroedinger_JPSi}) corresponds to $\Psi_{S}$ in~(\ref{eq:pNRQCDLagrangian})),
$P_{\mu}$ is the momentum of the moving quarkonium, $D_{\xi}=\partial_{\xi}+ig[A_{\xi},\Psi_{O}]$,
$A_{\mu}$ is the gauge field, and $V_{S}(r),\, V_{O}(s),\, V_{A}(r)$
and $V_{B}(r)$ are the potentials for singlet, octet diquarks and
a transition matrix elements. While in pQCD they are given by well-known
perturbative expressions, for a quarkonium propagating inside matter
they become more complicated and get an absorptive part. A detailed study of the absorption mechanism
is out of scope of this paper and will be presented elsewhere.

\section{Schroedinger equation from analyticity and unitarity}

\label{sec:AnalyticitySchroedinger}Notice that the single-gluon exchange
in $t$-channel gives only a Coulomb term in the effective potential.
As we have discussed in the introduction, such an approach can be
justified only for very heavy quark states (say bottomonium), whereas
for the $J/\Psi$ meson higher order corrections are essential. This
fact manifests itself in the developed phenomenology ~\cite{Digal:2005ht,Eichten:1979ms,Arafah:1983,Jacobs:1986,Gara:1990,Ikhdair:1992,Brambilla:1992,Fulcher:1993,Oh:2002,Blanck:2011}
in the rest frame, in which an additional confining potential is added
to one gluon exchange. 
This extra term is generated by nonperturbative interactions and introduced
into the Bethe-Salpeter either as additional vector or scalar $t$-channel
contribution (see e.g.~\cite{Resag:1995}).

Bearing this experience in mind, in this section we are going to obtain
the Schroedinger equation for the non-relativistic system without
using the explicit form of $\tilde{A}$ in~Eq.~(\ref{eq:BSE_def-1}).
In the following discussion, it is convenient to write the amplitude
$\chi$ as a function of the variables $s=\left(p_{-}-p_{+}\right)^{2}$and
$p=p_{-}+p_{+}$ instead of parton momenta $p_{\pm}$ . The amplitude
$\chi$ is analytic outside the real axis, so it can be represented
as 
\begin{equation}
\chi\left(p,\, M^{2}\right)=\frac{1}{2\pi}\int_{M^{2}}^{\infty}ds'\,\frac{\mbox{Im}\,\chi(s',\, p)}{s'-M^{2}+i0},\label{eq:disp_1}
\end{equation}
where $M$ is the mass of the $\bar{q}q$ meson. The imaginary part
of the amplitude $\chi$ can be evaluated directly from (\ref{eq:BSE_def-1}),
using the unitarity constraints 
\begin{equation}
\mbox{Im}\,\chi(s',\, p)=\int d^{3}q\,\chi(s',\, q)\, A_{2\to2}\left(s',\,\left(q-p\right)^{2}\right)\delta\left(s'-\left(2q-p\right)^{2}\right)+\sum_{n=3}^{\infty}\chi_{1\to n}\bigotimes A_{n\to2}\label{eq:disp_2}
\end{equation}
In this equation $A_{2\to2}\left(s',\, p-p'\right)$ is the 2-particle
scattering amplitude. If we neglect the contributions of higher Fock states
($n\geq3$ in Eq.~(\ref{eq:disp_2})) in the large-$m_{q}$ limit,
we can re-write Eq.~(\ref{eq:disp_1}) and Eq.~(\ref{eq:disp_2})
in the following form

\begin{equation}
\chi\left(p,\, M^{2}\right)=\frac{1}{2\pi}\int d^{3}q\int_{M^{2}}^{\infty}ds'\,\frac{\chi(s',\, q)}{s'-M^{2}+i0}\, A_{2\to2}\left(s',\,\left(q-p\right)^{2}\right)\delta\left(s'-\left(2q-p\right)^{2}\right)+\mathcal{O}\left(\frac{\epsilon}{m_{G}}\right),\label{eq:disp_3}
\end{equation}
where $m_{G}\approx$ 1 GeV is the effective mass of gluon~\cite{Graziani:1984cs}.
These estimates stem from the assumption that the second term in \eq{eq:disp_2}
has the same order of the magnitude as the first one, but its contribution
starts to be essential for $s'\,\geq\,(2m_{q}+m_{G})^{2}$. In the
dispersion integral the first term shows the enhancement $\propto1/(M\epsilon)$,
while the second has the denominator $\propto1/(M\, m_{g})$. It is
difficult to make better estimates for accuracy of our calculation
due to the lack of understanding of non-perturbative QCD.

It should be stressed that for very heavy quark-antiquark systems,
at small distances the omitted terms have an additional
suppression of the order of $\mathcal{O}(\alpha_{s}^{2}(m_{q}))$.
In the case of realistic system such as the $J/\Psi$-meson we cannot
use this smallness, and in order to evaluate the highly excited states in~\eq{eq:disp_2}
we have to rely on models. It turns out that in phenomenological
models for the $J/\Psi$-meson,  a substantial contribution stems from
the string-like potential. Indeed, $V_{{\rm eff}}(r_{J/\psi})-V_{{\rm pQCD}}(r_{J/\psi})\,\sim\,\sigma\langle r_{J/\psi}\rangle\,\approx\,0.4\, GeV$
for the string tension $\sigma=1\, GeV/fm$. Bearing this fact in
mind we can estimate the contribution of the multi-particle states
in the unitarity constraint, by comparing the mass of the next string
excitation or, in other words, the mass of the next resonance on the
$J/\Psi$ Reggeon trajectory with mass $M_{R}$. This contribution
is suppressed by the parameter

\begin{equation}
Q=\frac{2\, M\epsilon}{M_{R}^{2}-M^{2}}.
\end{equation}

The binding energy $\epsilon$ is a poorly defined object, since the mass
of the heavy quark is scheme-dependent. For the case of charm quark,
the estimates for the mass $m_{c}$ vary between 1.27~GeV and 1.8
GeV, so as an upper value estimate, we take $\epsilon_{{\rm max}}\approx2M_{D}-M_{J/\psi}\approx600\,{\rm MeV}$.
Taking $M=M_{J/\psi}$ and $M_{R}=M_{\psi(3686)}$~\cite{PDG:2014}
for the first excited state in a channel with $J/\psi$ quantum numbers,
we get an upper limit for the parameter $Q_{{\rm max}}\approx1/2$.
In the small-$\epsilon$ approximation we may simplify~(\ref{eq:disp_3})
as 
\begin{equation}
A_{2\to2}\left(s',\,\left(q-p\right)^{2}\right)\approx A_{2\to2}\left(M^{2},\,\left(q-p\right)^{2}\right).\label{ASUMP}
\end{equation}
It should be mentioned that the assumption of Eq.~(\ref{ASUMP}) means
that amplitude $A_{2\to2}\left(s',\,\left(q-p\right)^{2}\right)$
has no singularities related to the quark-antiquark state, and this
amplitude can be viewed as a sum of quark-antiquark irreducible Feynman
diagrams in QCD. Introducing a new function 
\begin{equation}
\Psi(s,\, q)=\frac{\chi(s,\, q)}{s-M^{2}+i0},
\end{equation}
we may rewrite~(\ref{eq:disp_3}) in the form 
\begin{equation}
\left(p^{2}-M^{2}\right)\Psi(M^{2},\, p)=\frac{1}{2\pi}\int d^{3}q\, V(q-p)\Psi(M^{2},\, p).\label{eq:LC_JPsi}
\end{equation}
In the instant form we may split the vector of relative motion $p$
into a part $p_{||}$ collinear to the quarkonium momentum $P$ and
part $\delta p$ orthogonal to it. After some trivial algebra, (\ref{eq:LC_JPsi})
reduces to a simple Schroedinger equation~(\ref{eq:Schroedinger_JPSi}).
In the front form~(\ref{eq:LC_JPsi}) reduces to the result obtained
in~\cite{Lepage:1980fj},

\begin{equation}
M^{2}\Psi=\left(\frac{p_{\perp}^{2}+m_{q}^{2}}{x(1-x)}+\hat{U}\right)\Psi,\,\label{eq:Relat}
\end{equation}
where $M$ is the mass of the meson, $q_{\perp}$ is the quark transverse
momentum, $x$ is the light-front fraction of the longitudinal momentum
of the quark, and $\hat{U}$ is the operator of the potential energy.
In the heavy quark mass limit, there are two natural small parameters,
$\mathcal{O}\left(\alpha_{s}\left(m_{q}\right)\right)$ and $\mathcal{O}\left(\Lambda_{{\rm QCD}}/m_{q}\right)$.
Typical momenta of the quarks in a nonrelativistic system are of order
$m_{q}\mathcal{O}\left(\alpha_{s}\left(m_{q}\right)\right)$, whereas
the binding energy is $\epsilon\sim m_{q}\mathcal{O}\left(\alpha_{s}^{2}\left(m_{q}\right)\right)$.
Taking into account that the light-front fraction $x\approx1/2+\delta x$
\footnote{Recall that fraction $x$ is invariant with respect to boost in
the longitudinal direction and can be calculated in the rest frame,
$x=(q_{0}+q_{z})/M=\frac{1}{2}+(\epsilon+\sqrt{m_{Q}\epsilon})/M$,
where $M$ is the mass of the quarkonium.} 
where $\delta x\approx\mathcal{O}\left(\alpha_{s}\left(m_{q}\right)\right)$,
we can see that~(\ref{eq:Relat}) reduces to 
\begin{equation}
\epsilon\Psi=\left(\frac{p_{\perp}^{2}}{m_{q}}+4m_{q}\,\delta x^{2}+\frac{1}{4m_{q}}\hat{U}\right)\Psi.
\end{equation}

If we make a Fourier transformation of $\Psi$ to coordinate space
according to 
\begin{equation}
\phi\left(\zeta,\, r_{\perp}\right)=\int\frac{d\delta x\, d^{2}q_{\perp}}{(2\pi)^{2}}\Psi\left(x=\frac{1}{2}+\delta x,\, q_{\perp}\right)e^{2im_{q}\zeta\,\delta x-iq_{\perp}\cdot r_{\perp}},
\end{equation}
where $\zeta$ is a parton separation along the $z^{-}$-axis,
and $\phi$ satisfies the ordinary Schroedinger equation 
\begin{equation}
\epsilon\phi\left(\zeta,\, r_{\perp}\right)=\left(\frac{-\Delta_{\perp}-\partial_{\zeta}^{2}}{m_{q}}+\hat{V}\right)\phi\left(\zeta,\, r_{\perp}\right),\label{eq:SchroedLC}
\end{equation}
where $\hat{V}=\hat{U}/(4m_{q})$. We would like to emphasize that
in contrast to~\cite{Gutsche:2014oua} the potential $\hat{V}$ is
local only in the coordinate space; in a LF frame the interaction
has a form of a convolution 
\begin{equation}
\hat{V}\phi\left(x,\, r_{\perp}\right)=\int V\left(x-x_{1},\, r_{\perp}\right)\phi\left(x_{1},\, r_{\perp}\right)dx_{1}.
\end{equation}
The Green function of the internal motion in the infinite momentum frame
satisfies an evolution equation 
\begin{eqnarray}
 &  & \frac{P^{+}}{M}\frac{\partial}{\partial x^{+}}\, G\left(x^{+},\zeta,\, r_{\perp};x_{1}^{+},\zeta_{1},r_{1,\perp}\right)=\label{eq:G_def}\\
 & = & \frac{-\Delta_{\perp}-\partial_{\zeta}^{2}+m_{q}^{2}}{m_{q}}G\left(x^{+},\zeta,\, r_{\perp};x_{1}^{+},\zeta_{1},r_{1,\perp}\right)+\hat{V}G\left(x^{+},\zeta,\, r_{\perp};x_{1}^{+},\zeta_{1},r_{1,\perp}\right),\nonumber 
\end{eqnarray}

Finally, we would like to address how the light front and rest frame
potentials are related to each other. From Eq.~(\ref{ASUMP}) we see
that this amplitude is a function of only four-dimensional transfered
momentum. As we can see from~(\ref{eq:Schroedinger_JPSi},\ref{eq:LC_JPsi}),
a relation between the potentials in different frames can be extracted
from the transformation of the gluon momentum $k=q-p$. In a rest
frame, the vector $k$ has a negligible $k_{0}$ component (instantaneous
approximation), so $k^{2}\approx-\vec{k}^{2}+\mathcal{O}\left(\alpha_{s}^{4}\right).$
In an infinite momentum frame, we may rewrite 
\begin{equation}
k^{2}=\left(q-p\right)^{2}=\left(x-x_{1}\right)\left(\frac{m_{q}^{2}+p_{\perp}^{2}}{x}-\frac{m_{q}^{2}+q_{\perp}^{2}}{x_{1}}\right)-q_{\perp}^{2}\approx-\frac{m_{q}^{2}}{x\, x_{1}}\left(x-x_{1}\right)^{2}-q_{\perp}^{2}+\mathcal{O}\left(\alpha_{s}^{3}\right),\label{gluonMomentumSquared}
\end{equation}
where light-front fractions are defined as $x=p^{+}/P^{+}$ and $x_{1}=q^{+}/P^{+}$.
Note that $x-x_{1}\sim\mathcal{O}\left(q_{\perp}/m_{q}\right)$, so
both terms are of the same order of smallness and the omission of
the terms in a previous line is justified due to extra $\sim\mathcal{O}\left(q_{\perp}/m_{q}\right)$
suppression. After some lengthy but straightforward evaluations (see
details in Appendix~\ref{sec:DerviationOfLCPotential}) we may get
for the interaction term in the infinite momentum frame 
\begin{eqnarray}
\left(\hat{U}\Psi\right)\left(x,\, r_{\perp}\right) & = & \int dx_{1}K\left(x,\, x_{1},\, r_{\perp}\right)\Psi\left(x_{1},\, r_{\perp}\right),\label{eq:U_def}
\end{eqnarray}
where the kernel

\begin{eqnarray}
K\left(x,x_{1},r_{\perp}\right) & = & \int_{-\infty}^{\infty}\frac{dz}{2\pi}\, V\left(\sqrt{r_{\perp}^{2}+z^{2}}\right)\exp\left(2im_{q}\left(x-x_{1}\right)\, z\right).\label{eq:K_def}
\end{eqnarray}

\begin{figure}
\includegraphics[scale=0.5]{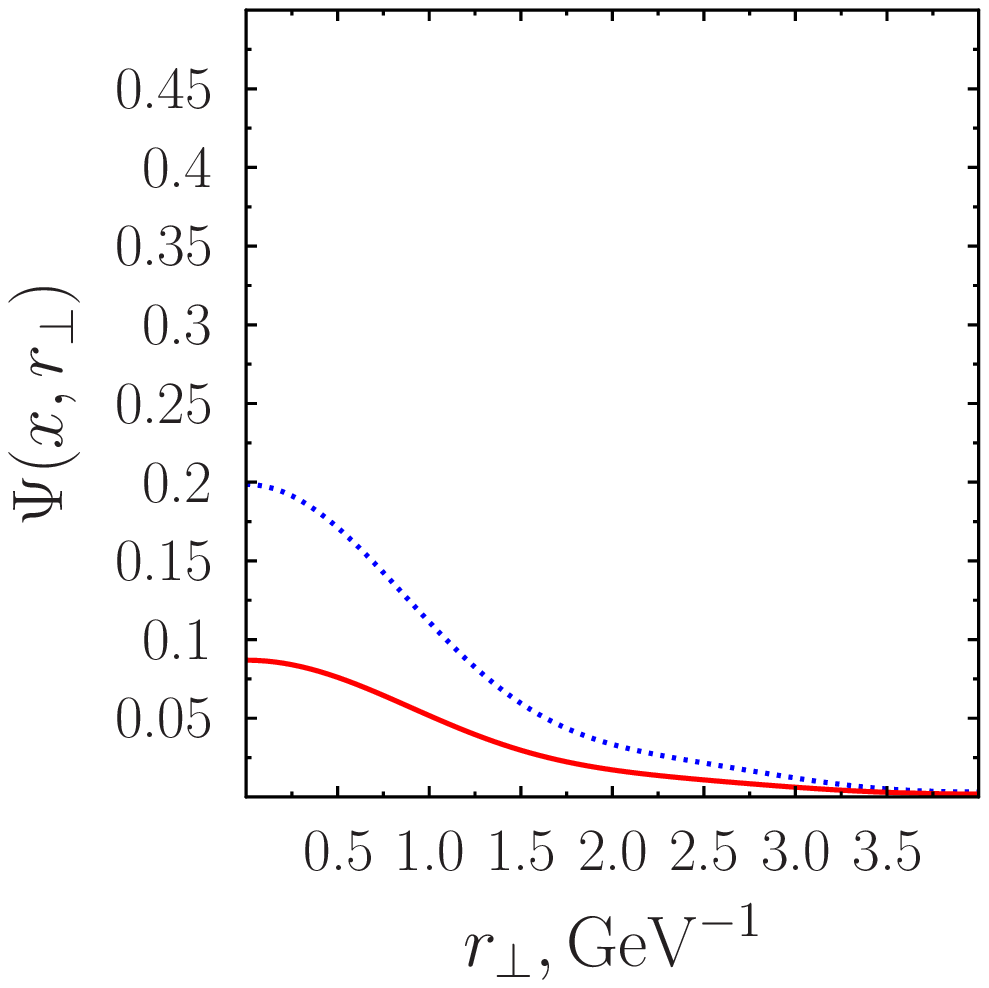}\includegraphics[scale=0.5]{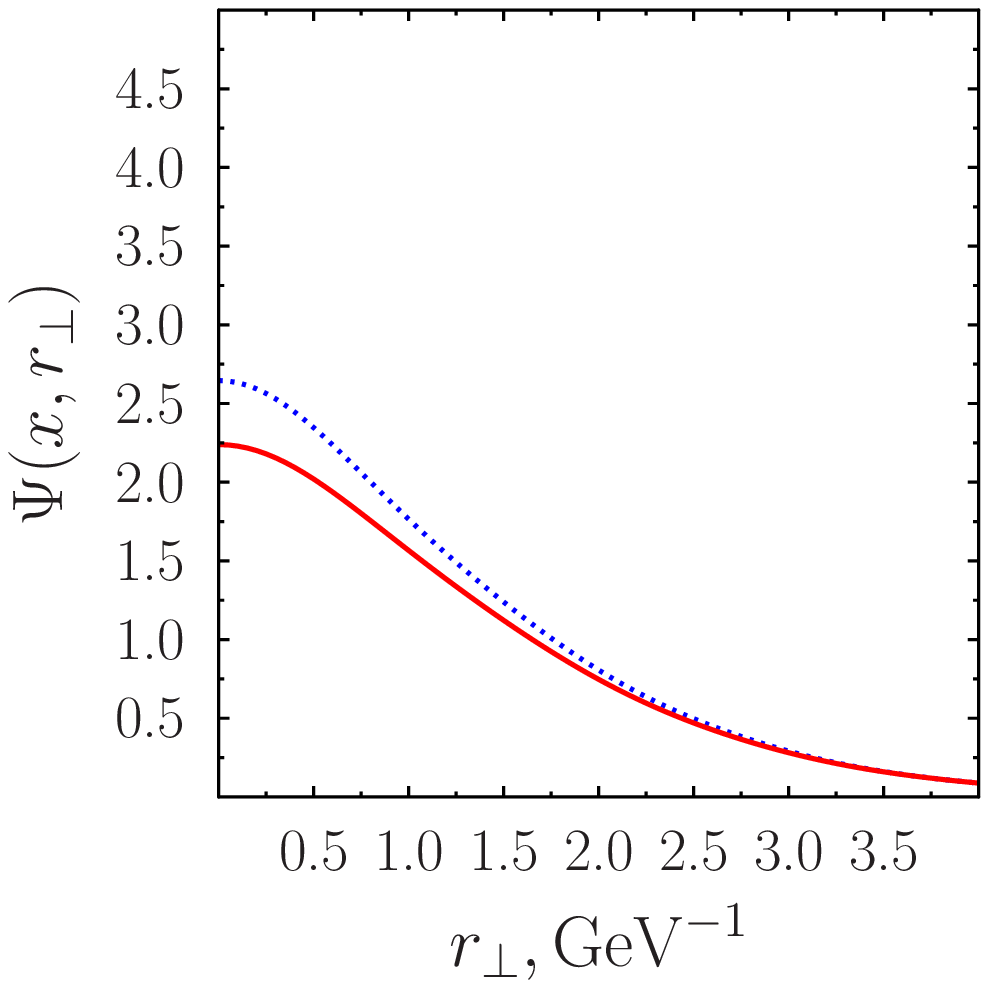}\includegraphics[scale=0.5]{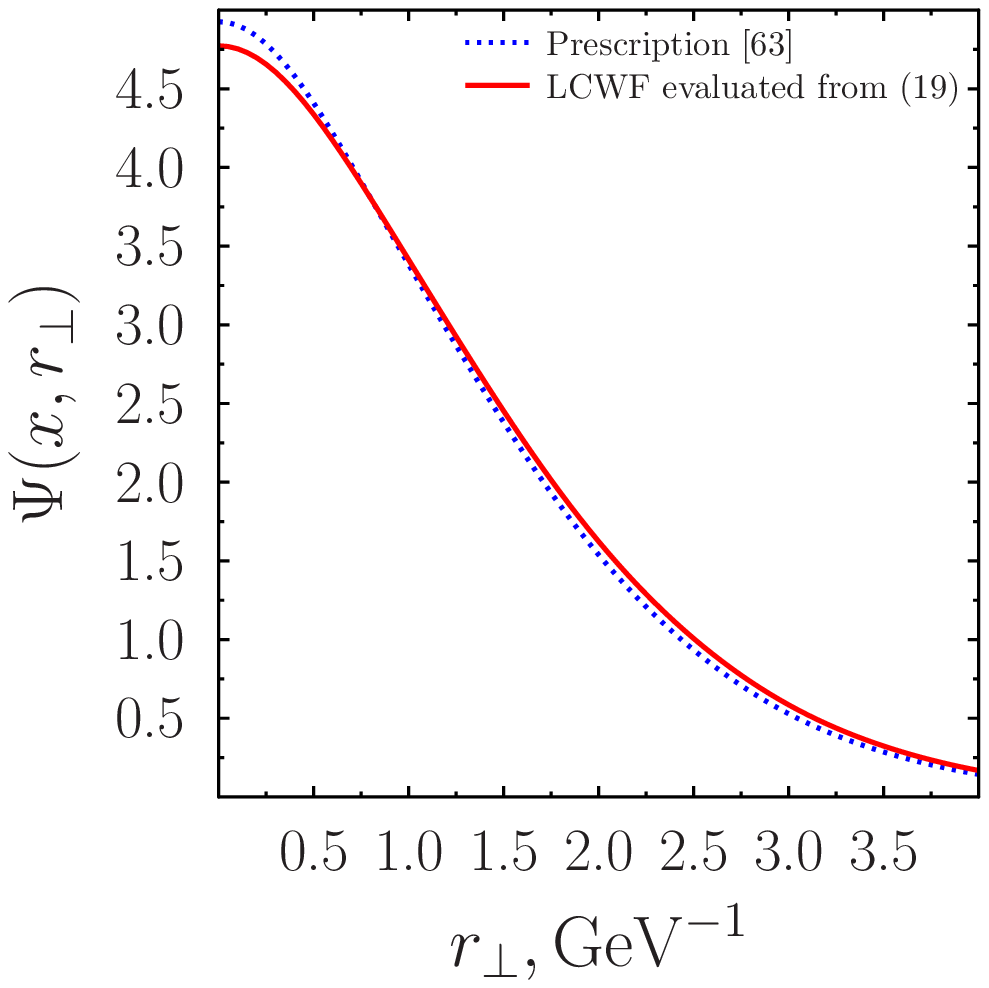}

\includegraphics[scale=0.5]{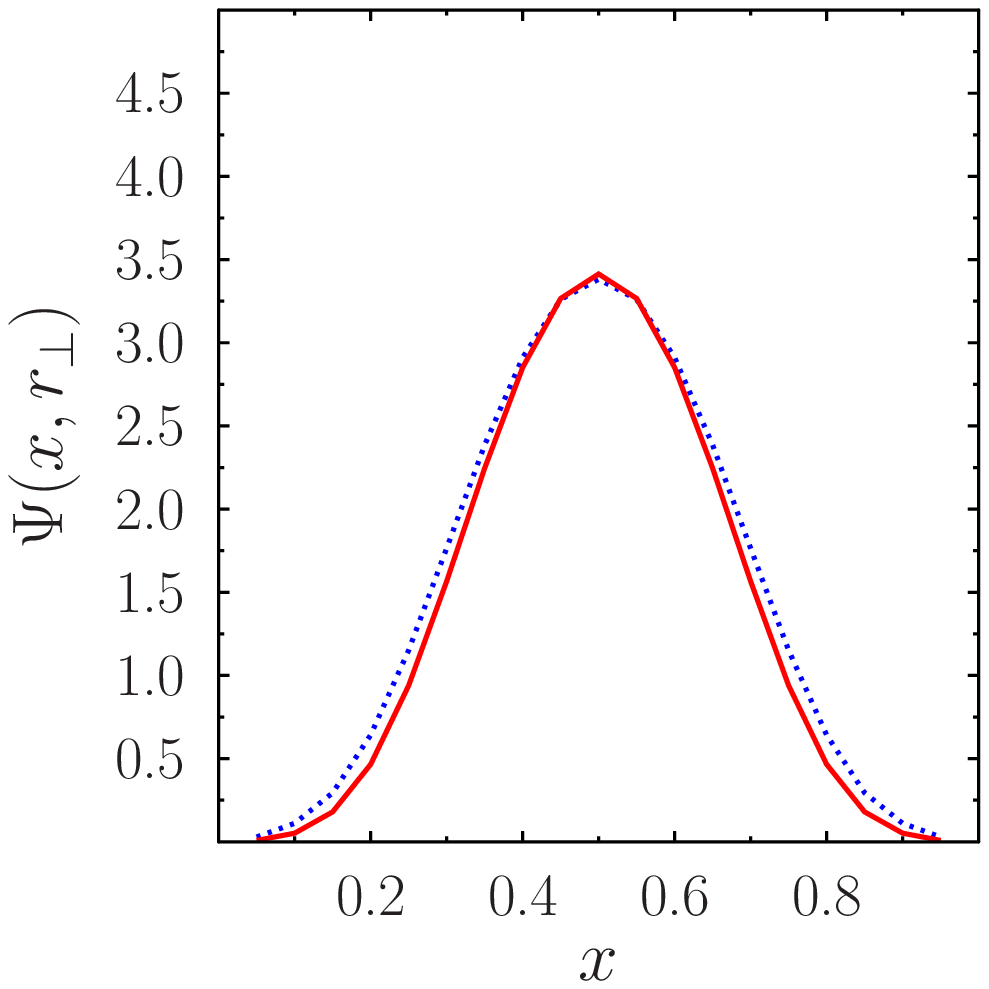}\includegraphics[scale=0.5]{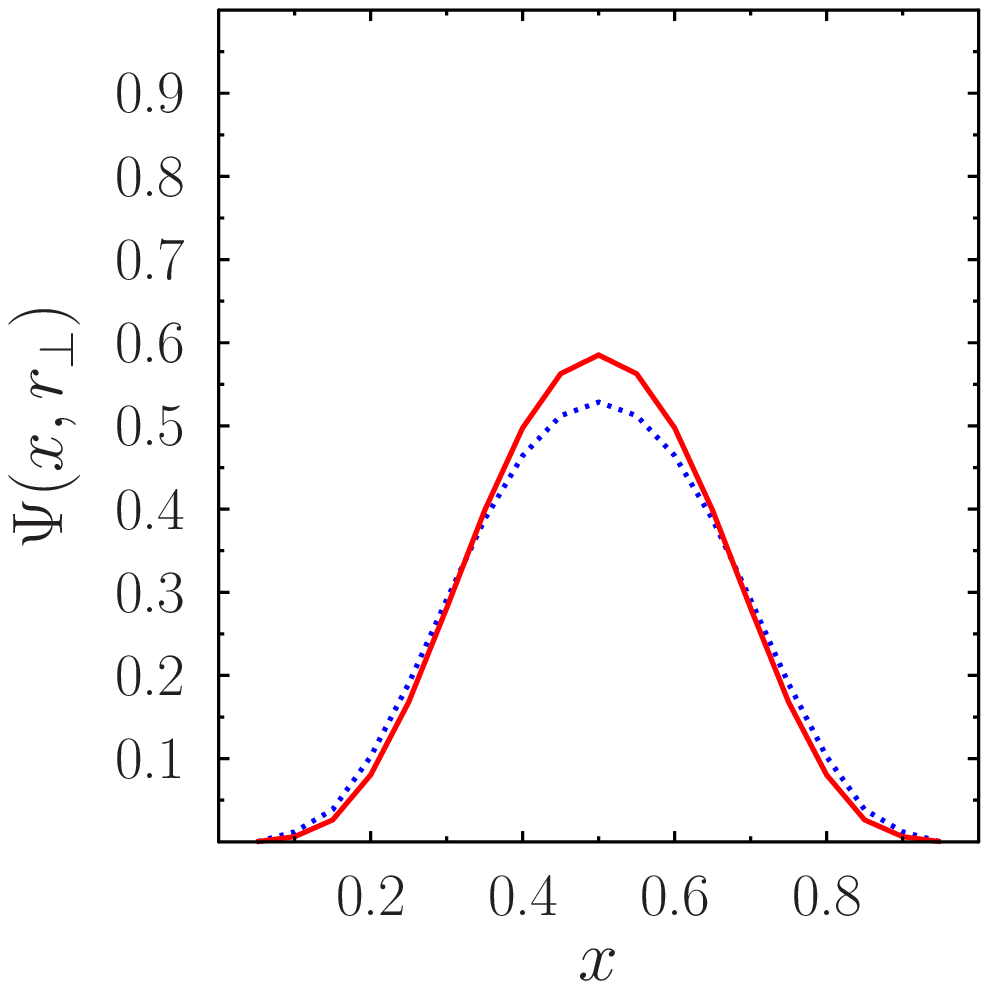}\includegraphics[scale=0.5]{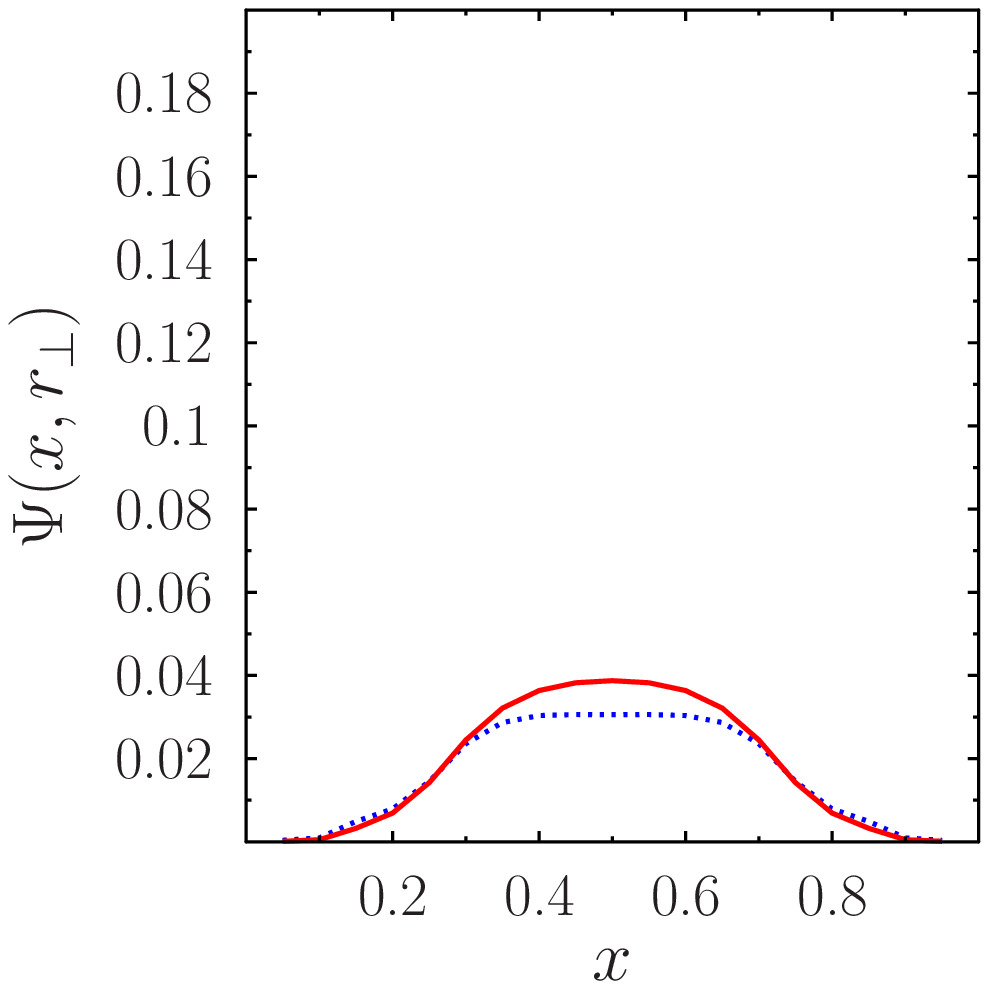}\caption{\label{fig:Comparison}Comparison of the ground state wave functions
evaluated with~(\ref{eq:G_def}) (solid red line) and with a prescription
from~\cite{Ivanov:2002kc} (dashed blue line). Upper row: $r_{\perp}$-dependence
at fixed~$x$. Lower row: $x$-dependence at fixed values of the parameter
$r_{\perp}$. }
\end{figure}

In Figure~\ref{fig:Comparison} we compare the ground state ($J/\psi$)
wave function evaluated from~(\ref{eq:G_def}) with the one obtained
with the phenomenological prescription~\cite{Ivanov:2002kc}. For
the sake of definiteness, in both cases we used a rest frame potential
from~\cite{Karsch:1987pv}. As we can see from the plots, for the
symmetric configuration ($x\approx0.5$) the wave function given by
a prescription~\cite{Ivanov:2002kc} coincides within the accuracy of
numerical evaluations with the eigenfunction of~(\ref{eq:G_def}).
However, for large-$x$ there is some disagreement between the two
wave functions. This could be easily understood, taking into account
that $x-1/2$ is a small $\mathcal{O}\left(\alpha_{s}\right)$ parameter
which is neglected. The average width of the charmonium wave function
is $\left\langle \left|x-\frac{1}{2}\right|\right\rangle \sim\langle p_{3}/2\, m_{q}\rangle\approx0.2$,
which signals that the relativistic effects, albeit small, are not
completely negligible.\\

\section{Conclusions}

In this paper we re-visit the problem of the wave function for the
 heavy quark-antiquark system in an arbitrary reference frame. In two
regions: (i) $m_{q}\,\alpha_{S}\left(m_{q}\right)\,\gg\Lambda_{QCD}$
and (ii) $\epsilon/\Lambda_{QCD}\,\ll\,1$, we constructed a Schroedinger
equation for the internal motion of $J/\psi$ moving in the laboratory
frame with relativistic center-of-mass momentum (see Eq.~(\ref{eq:Schroedinger_JPSi})).
In these cases, this equation reduces to the rest frame and infinite momentum
Schroedinger equations, and thus gives a smooth interpolation between
the two limits. We studied a relation between the rest frame and light-front
potentials of quarkonium. We found that the two are related by a linear
transformation~(\ref{eq:U_def}), which corresponds to a large-$m_{q}$
limit of a nonlinear transformation in~\cite{Trawinski:2014msa,Gutsche:2014oua}. The
omitted terms are of higher order in~$\mathcal{O}(\Lambda_{QCD}\Big{/}\left(\alpha_{s}\left(m_{q}\right)\right)$
and $\mathcal{O}\left(\epsilon\Big{/}\Lambda_{QCD}\right)$ and their
inclusion is not justified in the two-parton approximation. We checked
the phenomenological prescription~\cite{Ivanov:2002kc} and found
that in the region $x\approx0.5$, which is relevant for charmonium-related
problems, it is well justified. Also, we got that in the infinite momentum
frame a potential is given by a convolution~(\ref{eq:U_def},\ref{eq:K_def})
in light-front variable $x$, in contrast to what was found in~\cite{Gutsche:2014oua}.

Our results could be relevant for the description of the final state interactions
of a wide class of processes including quarkonium production on
nuclei or plasma. Besides, they could be extended to the description of
final state interactions in the production of weakly bound systems, such as
for example the deuteron.

\section*{Acknowledgments}

This work was supported in part by Fondecyt (Chile) grants No. 1130543,
1140390, 1140842 and 1140377. We are grateful to Valery E. Lyubovitskij
for discussion of potential transformation properties in the AdS/QCD
framework and for providing  the reference ~\cite{Gutsche:2014oua}.

\appendix

\section{Derivation of the Schroedinger equation}

\label{sec:BSE_derivation}The Bethe-Salpeter equation (BSE) has a
form 
\begin{equation}
i\chi_{ab}\left(p_{-},-p_{+}\right)=S_{ac}\left(p_{-}\right)\int\frac{d^{4}q}{(2\pi)^{4}}\tilde{A}_{ce,fd}\left(p_{-},q-p_{+}-p_{-},q-p_{+}\right)\chi_{ef}\left(q,\, q-p_{-}-p_{+}\right)S_{db}\left(-p_{+}\right),\label{eq:BSE_def}
\end{equation}
where $p_{-}$ and $-p_{+}$ are the momenta of the quark and antiquark
respectively, and we explicitly have shown lower Dirac/flavour indices.
The Bethe-Salpeter amplitude defined as 
\[
\chi_{ab}\left(p_{1},\, p_{2}\right)=\int d^{4}x_{1}d^{4}x_{2}e^{i\left(p_{1}\cdot x_{1}-p_{2}\cdot x_{2}\right)}\langle0|\psi_{a}\left(x_{1}\right)\bar{\psi}_{b}\left(x_{2}\right)|P\rangle
\]
is not a wave function, but can be projected to a wave function integrating
along some time-like curve. The propagator $S$ in leading order
may be replaced by the free quark propagator, and the amplitude $\tilde{A}$
is dominated by a single-gluon exchange in the $t$-channel. Note however
that single-gluon exchange produces just a Coulomb potential, whereas
in realistic models we have also a string, so we \emph{assume} that
the gluon propagator has a form $\Pi_{\mu\nu}=g_{\mu\nu}\tilde{V},$
where $\tilde{V}$ is a Fourier image of the potential. In this case
we may rewrite the equation~(\ref{eq:BSE_def}) as 
\begin{equation}
i\chi\left(p_{-},-p_{+}\right)=S\left(p_{-}\right)\int\frac{d^{4}q}{(2\pi)^{4}}\tilde{V}\left(q-p_{-}\right)\gamma_{\mu}\chi\left(q,\, q-p_{-}-p_{+}\right)\gamma_{\mu}S\left(-p_{+}\right),\label{eq:BSE_1g}
\end{equation}

Starting from this equation we omit the Dirac indices since we have
a matrix identity. 
If we denote $p_{0}=P_{J/\Psi}/2$ and define small deviations as
$\delta p_{-}=p_{-}-p_{0},\quad-\delta p_{+}=-p_{+}-p_{0}$, we may
expect that a deviation of quark momenta from $p_{0}$ are small;
the terms $\sim\delta p_{i}$ may be neglected in the numerator but should
be retained in the denominator, so we get  
\begin{equation}
S\left(p_{-}\right)=\frac{\hat{p}_{-}+m_{q}}{p_{-}^{2}-m_{q}^{2}+i0}\approx\frac{\hat{p}_{0}+m_{q}}{2m_{q}}\,\frac{1}{\frac{p_{0}\cdot\delta p_{-}}{m_{q}}-\frac{\delta\vec{p}_{-}^{2}}{2m_{q}}+i0}=\Lambda_{+}g\left(\delta p_{-}\right)
\end{equation}
\begin{equation}
S\left(-p_{+}\right)=\frac{-\hat{p}_{+}+m_{q}}{p_{+}^{2}-m_{q}^{2}+i0}\approx\frac{\hat{p}_{0}-m_{q}}{2m_{q}}\,\frac{1}{-\frac{p_{0}\cdot\delta p_{+}}{m_{q}}-\frac{\delta\vec{p}_{+}^{2}}{2m_{q}}+i0}=\Lambda_{-}g\left(-\delta p_{+}\right)
\end{equation}
where we introduced a shorthand notation for the projectors 
\[
\Lambda_{\pm}\equiv\frac{\hat{p}_{0}\pm m_{q}}{2m_{q}}
\]
and a new function 
\[
g\left(\delta p\right)=\frac{1}{\frac{p_{0}\cdot\delta p}{m_{q}}-\frac{\delta\vec{p}^{2}}{2m_{q}}+i0}
\]

Multiplying both parts of~(\ref{eq:BSE_1g}) by $\Lambda_{+}\gamma_{\nu}$
from the left and by $\gamma_{\nu}\Lambda_{-}$ from the right, after
some manipulations with Dirac algebra, we get 
\begin{eqnarray}
i\tilde{\chi}\left(p_{-},-p_{+}\right) & = & g\left(\delta p_{-}\right)g\left(-\delta p_{+}\right)\int\frac{d^{4}q}{(2\pi)^{4}}\tilde{V}\left(q-p_{-}\right)\tilde{\chi}\left(q,\, q-p_{-}-p_{+}\right),\label{eq:BSE_1g-1-1-1}\\
\tilde{\chi} & \equiv & \Lambda_{+}\gamma_{\mu}\chi\gamma_{\mu}\Lambda_{-}
\end{eqnarray}

In a moving system, a projection of the gluon momentum $q-p$ onto
the direction of vector $p_{0}$ is $\mathcal{O}\left(\alpha_{s}^{2}\right)$,
whereas all the other components are $\mathcal{O}\left(\alpha_{s}\right)$
so we may assume that the former can be neglected (it is a generalization of
the instantaneous approximation in QCD). Let us introduce a vector
\begin{equation}
n_{0}=\left(\sinh\eta,\cosh\eta,\vec{0}_{\perp}\right).
\end{equation}
where $\eta=\ln\left(P_{{\rm J/\psi}}^{+}/M_{{\rm J/\psi}}\right)$,
and make a Sudakov decomposition of the vector $q$, 
\begin{equation}
q=\lambda\frac{p_{0}}{m_{q}}+\tau n_{0}+q_{\perp}
\end{equation}
In a rest frame the projection of the vector $q$ onto $p_{0}$ is
suppressed as $\mathcal{O}\left(\alpha_{s}\right)$ compared to the
other components, for this reason in a moving frame we expect that
dependence on $\lambda$ should be negligible, \emph{i.e.} 
\begin{equation}
\tilde{V}=\tilde{V}\left(\tau,\, q_{\perp}\right)
\end{equation}

We define a wave function as 
\[
\Psi(\tau,\, q_{\perp};\, P)=\int d\lambda\,\tilde{\chi}\left(q,\, q-P\right),
\]
so~(\ref{eq:BSE_1g-1-1-1}) can be rewritten as 
\begin{equation}
i\tilde{\chi}\left(p_{-},-p_{+}\right)=g\left(\delta p_{-}\right)g\left(-\delta p_{+}\right)\int\frac{d\tau dq_{\perp}}{(2\pi)^{4}}\tilde{V}\left(\tau-\tau_{-},q_{\perp}-p_{\perp}\right)\Psi(\tau,\, q_{\perp};\, P)\label{eq:Schr_lc}
\end{equation}
and $\tau_{-}=p_{-}\cdot n_{0}$. Defining $\lambda_{-}=p_{-}\cdot p_{0}/m_{q}$
and taking the integral over it in both parts of~(\ref{eq:Schr_lc}),
we may get 
\begin{equation}
\left(E-\frac{\tau^{2}}{m_{q}}-\frac{p_{\perp}^{2}}{m_{q}}\right)\Psi=\int\frac{d\tau d^{2}q_{\perp}}{\left(2\pi\right)^{3}}\tilde{V}\left(\tau-\tau_{-},q_{\perp}-p_{\perp}\right)\Psi(\tau,\, q_{\perp};\, P)
\end{equation}
where $\tau=-p\cdot\bar{p}_{0}=-\sinh\eta\hat{p}_{0}+\cosh\eta\hat{p}_{3}$.
In coordinate space, the corresponding wave function is 
\[
\Psi\left(\zeta,\, r_{\perp}\right)=\int\frac{d\tau d^{2}q_{\perp}}{(2\pi)^{3}}e^{i\zeta\tau+ir_{\perp}\cdot q_{\perp}}\Psi(\tau,\, q_{\perp};\, P)
\]
where the parameter $\zeta$ is related to ordinary coordinates as 
\[
\zeta=x_{0}\sinh\eta-x_{3}\cosh\eta.
\]
and has a meaning of $z$-coordinate in the $J/\Psi$ rest frame. The
corresponding eigenvalue equation has a form of a Schroedinger equation
in the rest frame,

\begin{equation}
\left(E+\frac{\partial_{\zeta}^{2}}{m_{q}}-\frac{p_{\perp}^{2}}{m_{q}}\right)\Psi\left(\zeta,r_{\perp}\right)=V\left(\zeta,r_{\perp}\right)\Psi\left(\zeta,r_{\perp}\right),
\end{equation}
and thus guarantees a correct spectrum. For a nonstationary state,
we have to replace $E\to i\partial_{\xi}$, where 
\[
\xi=x_{3}\sinh\eta-x_{0}\cosh\eta,
\]
so the equation of motion is

\begin{equation}
\left(i\partial_{\xi}+\frac{\partial_{\zeta}^{2}}{m_{q}}-\frac{p_{\perp}^{2}}{m_{q}}\right)\Psi\left(\zeta,r_{\perp}\right)=V\left(\zeta,r_{\perp}\right)\Psi\left(\zeta,r_{\perp}\right),
\end{equation}
Replacing 
\begin{eqnarray*}
\partial_{\xi} & = & -\left(\cosh\eta\partial_{0}+\sinh\eta\partial_{3}\right),\\
\partial_{\zeta} & = & -\left(\sinh\eta\partial_{0}+\cosh\eta\partial_{3}\right),
\end{eqnarray*}
in the lab-frame, we end up with the second order (w.r.t. time) equation

\begin{equation}
\left(-i\left(\cosh\eta\partial_{0}+\sinh\eta\partial_{3}\right)+\frac{\left(\sinh\eta\partial_{0}+\cosh\eta\partial_{3}\right)^{2}}{m_{q}}-\frac{p_{\perp}^{2}}{m_{q}}\right)\Psi=V\left(\zeta,r_{\perp}\right)\Psi.
\end{equation}

\section{Derivation of~(\ref{eq:U_def},\ref{eq:K_def})}

\label{sec:DerviationOfLCPotential}As was discussed in section~\ref{sec:Comoving},
in order to find a relation between the rest frame and light front
potentials, it is necessary to rewrite the gluon momentum in terms
of the light-front components, as~(\ref{gluonMomentumSquared})

\begin{equation}
k^{2}=\left(q-p\right)^{2}=\left(x-x_{1}\right)\left(\frac{m_{q}^{2}+p_{\perp}^{2}}{x}-\frac{m_{q}^{2}+q_{\perp}^{2}}{x_{1}}\right)-q_{\perp}^{2}\approx-\frac{m_{q}^{2}}{x\, x_{1}}\left(x-x_{1}\right)^{2}-q_{\perp}^{2}+\mathcal{O}\left(\alpha_{s}^{3}\right).
\end{equation}
where for the sake of brevity we introduced a notation $k=q-p$ for
the gluon momentum. If $\Psi$ is the light-front wave function, and
$\tilde{\Psi}\left(x_{1},p_{\perp}\right)$ is its Fourier transform
over the transverse components, for the interaction term we get

\begin{eqnarray}
\hat{U}\Psi & = & \int\frac{d^{2}p_{\perp}}{(2\pi)^{2}}e^{ip_{\perp}\cdot r_{\perp}}\int\frac{dx_{1}}{2\pi}\frac{d^{2}k_{\perp}}{(2\pi)^{2}}\tilde{V}\left(k_{\perp}^{2}+\frac{m_{q}^{2}}{x\, x_{1}}\left(x-x_{1}\right)^{2}\right)\tilde{\Psi}\left(x_{1},p_{\perp}-k_{\perp}\right)=\\
 & = & \int\frac{d^{2}p_{\perp}}{(2\pi)^{2}}e^{ip_{\perp}\cdot r_{\perp}}\int\frac{dx_{1}}{2\pi}\frac{d^{2}k_{\perp}}{(2\pi)^{2}}d^{2}\rho_{\perp}e^{-i\left(p_{\perp}-k_{\perp}\right)\cdot\rho_{\perp}}\tilde{V}\left(k_{\perp}^{2}+\frac{m_{q}^{2}}{x\, x_{1}}\left(x-x_{1}\right)^{2}\right)\Psi\left(x_{1},\rho_{\perp}\right)=\nonumber \\
 & = & 2\int dx_{1}\frac{d^{2}k_{\perp}}{(2\pi)^{2}}r^{2}dr\,\frac{\sin\left(r\sqrt{k_{\perp}^{2}+\frac{m_{q}^{2}}{x\, x_{1}}\left(x-x_{1}\right)^{2}}\right)}{r\sqrt{k_{\perp}^{2}+\frac{m_{q}^{2}}{x\, x_{1}}\left(x-x_{1}\right)^{2}}}V(r)e^{-ik_{\perp}\cdot r_{\perp}}\Psi\left(x_{1},r_{\perp}\right)=\nonumber \\
 & = & \int dx_{1}K\left(x,x_{1},r_{\perp}\right)\Psi\left(x_{1},r_{\perp}\right),\nonumber 
\end{eqnarray}
where we introduced a kernel $K$ defined as

\begin{eqnarray}
K\left(x,x_{1},r_{\perp}\right) & = & 2\int\frac{k_{\perp}dk_{\perp}}{2\pi}dr\,\frac{r\,\sin\left(r\sqrt{k_{\perp}^{2}+\frac{m_{q}^{2}}{x\, x_{1}}\left(x-x_{1}\right)^{2}}\right)}{\sqrt{k_{\perp}^{2}+\frac{m_{q}^{2}}{x\, x_{1}}\left(x-x_{1}\right)^{2}}}V(r)J_{0}\left(k_{\perp}r_{\perp}\right)\label{eq:K_1}
\end{eqnarray}
In the nonrelativistic limit we may approximate $x\, x_{1}\approx1/4$
and simplify the kernel~(\ref{eq:K_1}) to 
\begin{equation}
K\left(x,x_{1},r_{\perp}\right)\approx2\int_{0}^{\infty}dr\, r\, V(r)\int_{0}^{\infty}\frac{k_{\perp}\, dk_{\perp}}{2\pi}\,\frac{\sin\left(r\sqrt{k_{\perp}^{2}+4m_{q}^{2}\left(x-x_{1}\right)^{2}}\right)}{\sqrt{k_{\perp}^{2}+4m_{q}^{2}\left(x-x_{1}\right)^{2}}}J_{0}\left(r_{\perp}k_{\perp}\right).
\end{equation}
The integral over $k_{\perp}$may be taken introducing a new variable
$\kappa=\sqrt{k_{\perp}^{2}+4m_{q}^{2}\left(x-x_{1}\right)^{2}}$
and using Integral $6.677.1$ from Ryzhik, Gradstein, as

\begin{eqnarray}
 &  & K\left(x,x_{1},r_{\perp}\right)\approx2\int_{0}^{\infty}dr\, r\, V(r)\int_{2m_{q}\left|x-x_{1}\right|}^{\infty}\frac{d\kappa_{\perp}}{2\pi}\,\sin\left(r\kappa\right)J_{0}\left(r_{\perp}\sqrt{\kappa^{2}-4m_{q}^{2}\left(x-x_{1}\right)^{2}}\right)=\\
 & = & 2\int_{0}^{\infty}dr\, r\, V(r)\times\frac{\theta\left(r>r_{\perp}\right)}{2\pi}\frac{\cos\left(2m_{q}\left|x-x_{1}\right|\sqrt{r^{2}-r_{\perp}^{2}}\right)}{\sqrt{r^{2}-r_{\perp}^{2}}}=\\
 & = & \frac{1}{\pi}\int_{r_{\perp}}^{\infty}dr\, r\, V(r)\frac{\cos\left(2m_{q}\left|x-x_{1}\right|\sqrt{r^{2}-r_{\perp}^{2}}\right)}{\sqrt{r^{2}-r_{\perp}^{2}}}.
\end{eqnarray}
Now again change a variable of integration, $z=\sqrt{r^{2}-r_{\perp}^{2}}$,
$dz=rdr/\sqrt{r^{2}-r_{\perp}^{2}}$, $z\in(0,\infty)$, $r=\sqrt{r_{\perp}^{2}+z^{2}}$,
so we get a result in a very clear and elegant form: 
\begin{eqnarray}
 & \Rightarrow & K\left(x,x_{1},r_{\perp}\right)=\frac{1}{\pi}\int_{0}^{\infty}dz\, V\left(\sqrt{r_{\perp}^{2}+z^{2}}\right)\cos\left(2m_{q}\left|x-x_{1}\right|z\right),\\
\left(\hat{U}\Psi\right)\left(x,r_{\perp}\right) & = & \int dx_{1}K\left(x,x_{1},r_{\perp}\right)\Psi\left(x_{1},r_{\perp}\right).
\end{eqnarray}
Notice that since $\cos(...)$ is an even function, we may omit an absolute
value sign, $|x-x_{1}|\to x-x_{1}$. Also, using symmetry w.r.t. $z\to-z$,
we can extend the integration region from $-\infty$ to $+\infty$,
adding an extra prefactor $1/2$, and also adding $i\int_{-\infty}^{+\infty}dz\sin\left(2m\left|x-x_{1}\right|z\right)$,
which after integration of a symmetric potential yields zero, so the
final result for the kernel $K$ may be cast into an equivalent form
\begin{eqnarray}
 &  & K\left(x,x_{1},r_{\perp}\right)=\int_{-\infty}^{\infty}\frac{dz}{2\pi}\, V\left(\sqrt{r_{\perp}^{2}+z^{2}}\right)\exp\left(2im_{q}\left(x-x_{1}\right)z\right).
\end{eqnarray}

\end{document}